
\input jnl

\preprintno{ILL--(TH)--92--8}

\title{Numerical Study of ${\rm c>1}$ Matter Coupled to
       Quantum Gravity}

\author{Simon M. Catterall}

\affil
Department of Applied Maths and Theoretical Physics
University of Cambridge
Silver St. Cambridge CB3 9EW
England

\author{John B. Kogut}

\affil
Loomis Laboratory of Physics
University of Illinois at Urbana--Champaign
1110 West Green Street
Urbana, Illinois 61801

\author{Ray L. Renken}

\affil
Department of Physics
University of Central Florida
Orlando, Florida 32816

\abstract{We present the results of a numerical simulation aimed
at understanding the nature of the `$\rm c=1$ barrier' in two
dimensional quantum gravity. We study multiple Ising models living
on dynamical $\phi^3$ graphs and analyse the behaviour of moments
of the graph loop distribution. We notice a universality
at work as the average properties of typical graphs from
the ensemble are determined only by the central charge.
We further argue that the qualitative
nature of these results can be understood from considering the effect
of fluctuations about a mean field solution in the Ising sector. }

\doublespace
\pageno=2
In recent years considerable progress has been made in understanding
the nature of some simple systems encorporating two dimensional quantum
gravity. Two approaches have been followed which are at first sight quite
different. The first of these employs techniques
borrowed from conformal field theory
to calculate the spectrum of anomalous dimensions in several
simple models [1-2]. In the second method a regularisation of the
continuum functional integrals is made in terms of random triangulations [3-5].
Where it has been possible to solve the models analytically the two
approaches have been in complete agreement. However, in both cases all
attempts to solve the models when the matter central charge $c$ becomes greater
than unity have failed. The lattice regularisation however has the
advantage that it allows these `strongly coupled' regions to be
explored using computer simulation techniques.

In this letter we present the results of one such study aimed at examining
whether the phase transition predicted for $c>1$ is a genuine physical
phenomena or merely an artefact of our current methods of solution [6].
A single Ising model, at criticality, can be shown to possess $c=1/2$, so
a set of $P$ noninteracting systems would (at least naively) correspond
to a central charge $c=P/2$. By tuning $P$ from one through three we can
search for any sign of discontinuous behaviour associated with such a
transition.
The partition function we study can be written as
$$Z\left(\beta, P\right)=\sum_{G_N}e^{PF\left(\beta, G\right)}$$
$$e^{F\left(\beta, G\right)}=\sum_{\sigma_i}e^{\beta G_{ij}\sigma_i\sigma_j}
\eqno(1)$$
The sum over $\phi^3$ graphs $G_N$ is restricted to those with $N$ nodes
and fixed spherical topology. The connectivity matrix $G_{ij}$ can be
taken to be unity when the sites $i$ and $j$ are neighbour on the
graph and zero otherwise.

We sample the space of all graphs $G_N$ by random moves which are
dual to direct lattice link flips as
described in [7], whilst cluster algorithms are used to provide an
efficient updating procedure for the Ising sector [8]. On a fixed graph
this partition function would simply factorise into $P$ copies of a single
Ising partition function, but one expects that the use of a dynamical lattice
induces an effective coupling between different Ising \lq species\rq.
To this end we collected Monte Carlo data for the standard Ising
observables such as the specific heat $C$, susceptibility $\chi$, and
generalised spin-spin correlation matrix $C^{\alpha\beta}\left(r\right)$

$$C^{\alpha\beta}\left(r\right)= \left\langle \sigma^\alpha
\left(0\right)\sigma^\beta\left(r\right)\right\rangle_C\eqno(2)$$
The indices $\alpha$ and $\beta$ refer to the species and $r$
measures the geodesic distance on the graph between the
spins. The latter quantity,
in particular, could be expected to show signs of this induced coupling by
the
occurrence of non-vanishing off diagonal entries. However, in spite of
relatively good statistics ($5\times 10^5$ sweeps for a range of $\beta$
from $0.3$ to $1.2$) we were unable (within the statistical
errors) to extract any reliable signal
that would indicate behaviour
inconsistent with that of a single Ising model.
As a sample of our data we have listed in table 1. the
effective masses extracted by simple exponential fits to the
diagonal components of the correlation function both for $P=1$ and
$P=3$. Except for a possible small additive renormalisation there
is no evidence for a qualitative difference between the two cases.
There was no observable signal in the
off diagonal components. It is possible that our
lattices are too small ($N=1000$ and $N=2000$), but this we feel is
unlikely since previous calculations for $P=1$ with this model (ref. [9-10])
have indicated the onset of scaling behaviour on systems of
this size. We thus conclude that the magnitude of any such coupling
in the effective spin action is indeed very weak.

It has been argued [6], that the coupling of $c>1$ matter to two
dimensional gravity creates instabilities in the worldsheet metric so
to establish an order parameter for such a transition we should
consider the effective gravitational partition function having traced over
the spin variables. Associated with this is the question of what
purely gravitational observables we should concentrate on. In our
discrete model the geometry is determined by
the local structure of the $\phi^3$ graph.  This in turn is reflected
in the size and frequency of its loops. A large loop corresponds to
a highly coordinated vertex in the direct lattice and hence large
scalar curvature.
Thus, a non trivial quantity to focus on is the
probability distribution $Q\left(l\right)$ for the lengths of loops $l$
in the graph. Moments
of this distribution are then simply related to expectation values
of moments of the scalar curvature density on the direct lattice.
Thus we are led to consider observables $r^k\left(\beta,P\right)$
defined by
$$r^k\left(\beta, P\right)=\sum_{l}Q\left(l\right)\left(1-l/6\right)^k,
\qquad l=3,\ldots ,N\eqno(3)$$
One might expect that the large $k$ moments are more sensitive to the
occurrence of large (macroscopic) loops in the lattice.
Fig. 1 shows a plot of the $k=2$ moment describing the fluctuations
in the local scalar curvature density as a function of $\beta$ for
$P=0$ (pure gravity), $P=1$, $P=2$ and $P=3$ using $N=2000$. On the figure
we also mark the result for a $D=1$ pure gaussian model on the same lattice.
The strong coincidence of the gaussian model result with the two
spin curve suggests a large universality at work, the local structure
of the graph appears to be insensitive to all but the central charge
of the matter and not the details of its fields or action.
Clearly the addition of Ising spins enhances the fluctuations in the local
curvature peaking strongly at large $P$ in the vicinity of the Ising
critical point ($\beta_c=0.77332..$). Furthermore the height of this
peak scales linearly with $P$ and hence the central charge. This is in
qualitative agreement with the results of [11]. We have checked that
this observation remains essentially the same if a different measure is
adopted for the sum over graphs (e.g the conformal weight determined
by adding a term $-{c\over 2}\sum_i \ln\left(l_i\right)$ to the action).
However there is nothing in this plot signaling any pathology as $P$
increases through two. These observations support a recent matrix model
study by Brezin et al. [12].

One possible explanation for this stems from the observation that any
$c\ge 1$ model is expected to have an infinite number of relevant
operators. Any lattice representation of the theory would, in principle,
have to include lattice
versions of these operators, whose couplings would all have to be
tuned to certain values for the system to become critical [13]. Since it
is not clear how to construct these lattice operators (a generalised
Jordan Wigner construction ?) and even less how to truncate the set
to a manageable number whilst preserving the central charge, we chose
to work with our simpler model. Since our susceptibility data when $P>1$ are
still consistent with the scaling expected at a continuous phase
transition,
any continuum theory defined in its vicinity is still a
candidate for a $c>1$ gravity coupled
theory.

For the case of pure gravity, the behaviour of $Q\left(l\right)$ at
large $l$ is well known [3].
$$Q\left(l\right)\sim e^{-\alpha_0l},\qquad \alpha_0=0.288..\eqno(4)$$
If we make the plausible assumption that this holds for $P>0$, and consider
the case of large $k$, it is easy to show that
$$\left\langle r^k\left(\beta, P\right)\right\rangle\sim {k! \over 6^k}\left(
{1\over \alpha\left(\beta, P\right)}\right)^k\eqno(5)$$
Hence, if we divide the Ising moments by the corresponding pure
gravity ones and consider the quantities
$$M_k\left(\beta, P\right)=\ln{{\left\langle r^k_P\right\rangle\over
\left\langle
r^k_0\right\rangle}}\eqno(6)$$
We expect
$$M_k\left(\beta, P\right)\sim k\ln{\left({\alpha\left(\beta, 0\right)\over
\alpha\left(\beta, P\right)}\right)}\eqno(7)$$
Thus the parameters of the asymptotic distribution can be probed by
fitting $M$ to $k$
as both $\beta$ and $P$ are varied.

In fig. 2, we plot $M_k\left(\beta=0.7,P=3\right)$ as a function of $k$ for
an $N=2000$ node graph. Clearly we see good evidence for the expected linear
behaviour, a least squares fit yielding a value of $\ln\left(\alpha_0/\alpha_3
\right)=0.18(1),\,\chi^2=0.1$. Using these fits we can compare the gradients
$\ln\left(\alpha_0/\alpha_P\right)$ for $P=1$, $P=2$ and $P=3$
from the $N=2000$ node
ensemble
with $\beta$ in the range $0.3-1.2$ (figs. 3,4,5.).
The $P=3$ and $P=2$  curves
now
show a prominent peak close to the critical coupling, indicating a rapid
decrease in the associated $\alpha_P$, whilst the single Ising model exhibits
a much weaker structure consistent with a broad plateau in the same region
of coupling constant. Furthermore the peak height increases from $0.17(2)$ to
$0.18(2)$ for three Ising models on going from $N=1000$ nodes to $N=2000$,
and from $0.09(2)$ to $0.13(2)$ for $P=2$. This is in contrast to
the single spin case where the peak falls from $0.12(4)$ to $0.07(2)$ on
going to the larger lattice. Whilst our errors are clearly large it is
possible that our results are hinting at a qualitative
difference in the thermodynamic behaviour of $P<1$ and $P>1$. Clearly
high statistics runs on larger lattices will be needed to
resolve this issue unambiguously.
One possible scenario would be that the exponential behaviour in the
tail of $Q\left(l\right)$ for $P\le1$ gets replaced by a power law
for $P>1$. This would then ensure that sufficiently high moments of
the scalar curvature would formally diverge signaling a breakdown of
the model. However at present such ideas are merely speculation
of how any possible continuum instability might manifest itself
in the lattice models.

It is perhaps surprising that the effect of adding Ising spins on the
intrinsic geometry is rather similar to that of gaussian scalar fields.
In the latter case the effective gravitational partition function
is given by
$$Z^\prime=\sum_{G_N} {\rm det}^{-{D\over2}}\left(-\Delta^G\right)\eqno(8)$$
Here $D$ counts the number of bosons (embedding dimension) and
the scalar Laplacian $\Delta^G_{ij}=G_{ij}-q\delta_{ij}$.
Using this representation it has been shown that branched polymers
(in the sense of the intrinsic geometry) dominate in the $D\to\infty$ limit.

In contrast, the critical region of the single Ising model
is usually represented in
terms of a Majorana fermion. On integration over this field one
would have expected to find a positive power of an appropriate
fermion determinant. Thus the dependence on graph $G_N$ could
be expected to be quite different from the bosonic case which is
not what we see.

We can gain some insight into this situation by
considering the following (standard) representation of the Ising free
energy $F\left(\beta, G\right)$ (eqn. 1) in terms of a
continuous scalar
field $\phi_i$.

$$e^F=2^{N}{\rm det}^{-{1\over 2}}\left({\beta\over 2} G\right)
\int D\phi e^{-{1\over 2\beta}\phi_i G^{-1}_{ij}
\phi_j +\sum_i\ln\cosh\phi_i}\eqno(9)$$

If the dependence of $F$ on graph $G_N$ could be evaluated exactly
we could determine which graphs dominated the
effective gravitational partition function in the Ising case.
Unfortunately an exact solution is not possible and we
may only compute $F\left(\beta, G\right)$ in
some approximate way.
The one loop approximation corresponds to expanding the field
$\phi$ about some stationary configuration and integrating over
small fluctuations.
The stationary configuration is determined by the usual
mean field equation
$$G^{-1}_{ij}\phi_j=\beta\tanh\left(\phi_i\right)\eqno(10)$$
Although the generic graph $G_N$ is random it still possesses a constant
coordination number $q=3$, so it is consistent to seek a homogeneous
solution to this equation
$$\phi^0/{\beta q}=\tanh\phi^0\eqno(11)$$
Now expanding the action about this stationary value $\phi^0$
to quadratic order and doing the resulting gaussian integrals leads
to an expression for the free energy to one loop of the form
(notice that the ill-defined ${\rm det}\left(G\right)$ piece has
cancelled)
$$F=F_1+F_2$$
$$F_1=N\left(\ln\cosh\phi^0-{\beta q\over 2}\tanh^2\phi^0\right)$$
$$F_2=-{1\over 2}{\rm Tr}\ln\left(1-{\beta\over \cosh^2\phi^0}G\right)
\eqno(12)$$
Notice that the mean field piece $F_1$ does not depend on graph $G$.
In the limit $\beta\to 0$ $F_2$ and $F_1$ vanish and we recover the
pure gravity result. Conversely, as $\beta\to\infty$, $F_2$ gives
no contribution whilst $F_1$ contributes the ($G$ independent)
expected piece $e^{-\beta Nq/2}$.
For general $\beta$ we can recast the $F_2$ term as
$$F_2=-{1\over 2}{\rm Tr}\ln\left(m^2-\Delta^G\right)\eqno(13)$$
where $\Delta^G$ is the scalar Laplacian on the graph and $m^2$ is given
by
$$m^2={\cosh^2\phi^0\over\beta}-q$$
Thus at $\beta=\beta_c={1/q}$ the mass $m$ vanishes and a massless
{\it bosonic} field emerges.
$$e^{F_1+F_2}\sim {\rm det}^{-{1\over 2}}\left(-\Delta^G\right)\eqno(14)$$
In this situation the number of Ising models $P$ plays
a role analogous to the dimension $D$ in the gaussian case and branched
polymers dominate for large $P$. However, since we expect
fluctuation effects to be very important for two dimensions in
determining the exact form of the effective action, the most we can
hope for our gaussian approximation is that it indicates
the sort of systematic trend we should expect. We feel that it
is at least consistent with our numerical results.

In conclusion, we have presented results from a simulation study
of noninteracting Ising models on a dynamical random lattice. This
was motivated by a desire to try to derive a better understanding of the
coupling of $c\ge 1$ matter to two dimensional quantum gravity.
Whilst we find no strong signal for a breakdown of the model in this
region, we have perhaps pinpointed the relevant observables which
might provide an unequivocal resolution of the question given a
larger scale study.

We have discussed some of the problems associated with a lattice formulation
of the problem, and demonstrated within the context of a loop expansion
an interpretation of at least some of the systematic trends we see
in our Monte Carlo data.

This work was supported, in part, by NSF grant PHY 87-01775 and the
numerical calculations were performed using the Florida State University
CRAY YMP. Also, we acknowledge National Science Foundation Support
through the Materials Research Laboratory at the University of
Illinois, Urbana-Champaign, grant NSF-DMR-20538. SMC would
like to acknowledge fruitful discussions with Ian Drummond and Sumit
Das.

\references

[1]  V. Knizhnik, A. Polyakov and A. Zamolodchikov, Mod. Phys. Lett.A3 819
1988.

[2]  J. Distler and H. Kawai, \np B321, 509, 1989.

[3]  J. Ambjorn, B. Durhuus and J. Frohlich, \np B257, 433, 1985.

[4]  F. David, \np B257, 543, 1985.

[5]  B. Boulatov, V. Kazakov, I. Kostov and A. Migdal, \np B275, 641, 1986.

[6]  N. Seiberg, Rutgers preprint RU-90-29.

[7]  S. Catterall, J. Kogut and R. Renken, \np B366, 647, 1991.

[8]  R. Swendson and J. Wang, \prl 58, 86, 1987.

[9]  S. Catterall, J. Kogut and R. Renken, \prd 45, 2957, 1992.

[10] J. Jurkiewicz, A. Krzywicki, B. Petersson and B. Soderberg, \pl B213, 511,
     1988.

[11] C. Baillie and D. Johnston, Colorado preprint COLO-HEP-276.

[12] E. Brezin and S. Hikami, \pl B283, 203, 1992.

[13] Sumit Das - private communication.

\endreferences

\figurecaptions

[1] $k=2$ moment of the loop distribution for $N=2000$ versus $\beta$.
    $P=1$ by $\times$, $P=2$ by $\diamond$, and $P=3$ by $\circ$.
    Solid lines indicate pure gravity ($P=0$) result together with
    the $D=1$ gaussian model result.

[2] $M_k\left(\beta=0.7,P=3\right)$ versus $k$, $N=2000$.

[3] $\ln\left(\alpha_0\over\alpha_1\right)$ versus $\beta$, $N=2000$.

[4] $\ln\left(\alpha_0\over\alpha_2\right)$ versus $\beta$, $N=2000$.

[5] $\ln\left(\alpha_0\over\alpha_3\right)$ versus $\beta$, $N=2000$.

\endfigurecaptions

\setbox\strutbox=\hbox{\vrule height11.5pt depth5.5pt width0pt}
$$\vbox{\tabskip=0pt \offinterlineskip
\halign to423pt{\strut#& \vrule#\tabskip=1em plus2em& \hfil#& \vrule#&
\hfil#\hfil& \vrule#& \hfil#& \vrule#
\tabskip=0pt\cr \noalign{\hrule}
& & \omit\hidewidth $\beta$ \hidewidth
& & \omit\hidewidth $m_{P=1}$ \hidewidth
& & \omit\hidewidth $m_{P=3}$ \hidewidth
& \cr \noalign{\hrule}
&& 0.50 && 0.520(4) && 0.530(1) & \cr\noalign{\hrule}
&& 0.60 && 0.334(2) && 0.338(1) & \cr\noalign{\hrule}
&& 0.65 && 0.245(2) && 0.247(1) & \cr\noalign{\hrule}
&& 0.70 && 0.151(1) && 0.162(1) & \cr\noalign{\hrule}
&& 0.75 && 0.069(1) && 0.089(1) & \cr\noalign{\hrule}
&& 0.80 && 0.029(1) && 0.035(1) & \cr\noalign{\hrule}
}}$$
Table 1. Masses obtained from least square fits to simple
exponentials, the fitting windows (which are the same for $P=1$ and
$P=3$) being determined from
the effective mass plots.
\endit